\documentclass{PoS}

\title{Cosmic ray penetration in diffuse clouds }

\ShortTitle{Cosmic ray penetration in diffuse clouds }

\author{\speaker{Giovanni Morlino}\\
        INFN -- Gran Sasso Science Institute, viale F. Crispi 7, 67100 L'Aquila, Italy\\
        INAF -- Osservatorio Astrofisico di Arcetri, Largo E. Fermi 5, Firenze, Italy \\
        E-mail: \email{giovanni.morlino@infn.gssi.it}}

\author{Stefano Gabici\\
        Astroparticule et Cosmologie (APC), CNRS, Universit\'e Paris 7 Denis Diderot, Paris, France\\
        E-mail: \email{gabici@apc.in2p3.fr}}
        
\author{Julian Krause\\
        Astroparticule et Cosmologie (APC), CNRS, Universit\'e Paris 7 Denis Diderot, Paris, France\\
        E-mail: \email{julian.bautista@apc.univ-paris7.fr}}        

\abstract{Cosmic rays are a fundamental source of ionization for molecular and diffuse clouds, influencing their chemical, thermal, and dynamical evolution. The amount of cosmic rays inside a cloud also determines the $\gamma$-ray flux produced by hadronic collisions between cosmic rays and cloud material.
We study the spectrum of cosmic rays inside and outside of a diffuse cloud, by solving the stationary transport equation for  cosmic rays including diffusion, advection and energy losses due to ionization of neutral hydrogen atoms. We found that the cosmic ray spectrum inside a diffuse cloud differs from the one in the interstellar medium (ISM) for energies smaller than $E_{br}\approx 100$ MeV, irrespective of the model details. Below $E_{br}$, the spectrum is harder (softer) than that in the ISM if the latter is a power law $\propto p^{-s}$ with $s$ larger (smaller) than $\sim0.42$.  As a consequence also the ionization rate due to CRs is strongly affected.
Assuming an average Galactic spectrum similar to the one inferred from AMS-2 and Voyager 1 data, we discuss the resulting ionization rate in a typical diffuse cloud.}

\FullConference{The 34th International Cosmic Ray Conference,\\
		30 July- 6 August, 2015\\
		The Hague, The Netherlands}

\begin{document}

\section{Introduction}
 \label{sec:intro}
The amount of penetration of cosmic rays (CRs) into molecular clouds (MCs) regulates the ionization level of clouds and dense cores (for reviews see \cite{dalgarno,ceccarelli,indriolo}) and thus affects their dynamical evolution and the process of star formation. Moreover, the exclusion of CRs from MCs can reduce their gamma-ray emission \cite{Skill-Strong76,gabici07}, which results from the decay of neutral pions produced in inelastic interactions of CRs in the dense gas (see \cite{mereview} for a review). Therefore, it is of prime importance to understand wether CRs do penetrate or not MCs.

The difficulty of modeling the problem of CR penetration into clouds resides in its highly non-linear nature: CRs generate magnetic turbulence at the cloud border due to streaming instability. The level of the turbulence in turn determines the diffusion coefficient of CRs and thus, presumably, their capability of penetrate clouds. Some discrepancy exists in the literature, different theoretical approaches to the problem giving different results. According to early papers, CRs with energies below tens or hundreds of MeV are effectively excluded from MCs \cite{Skill-Strong76,Ces-Voelk78}, while in a more recent work, \cite{EZ11} found out that the CR intensity is only slightly reduced inside clouds. However, a direct comparison of the two approaches is not straightforward, since the former are kinetic approaches, while the latter a two fluid ones. 
An implicit assumption in all these papers is the fact that streaming instability would enhance the magnetic turbulence and cause the exclusion of CRs from clouds.

Here, we present a solution of the steady-state kinetic transport equation of CRs along a magnetic flux tube that encompasses a MC. 
We generalize the simplified two-zones (in- and out-side of the cloud) kinetic approaches by \cite{Skill-Strong76} and \cite{Ces-Voelk78} by considering the full spatially dependent equation.
Remarkably, we find that the exclusion of CRs from diffuse clouds of typical column density $N_H \approx 3 \times 10^{21}$~cm$^{-2}$ is effective below an energy of $\approx 100$~MeV, independently on the presence or not of streaming instability. In fact, the exclusion energy $E_{br}$ depends only (and quite weakly) on the physical parameters that characterize the ISM and the gas in the cloud. This result suggests that: {\it i)} the suppression of the gamma-ray emission from a cloud due to CR exclusion is not significant (the threshold for neutral pion production being $\approx 280$~MeV), and {\it ii)} the intensity of CRs is suppressed inside MCs at the particle energies which are most relevant for ionization.

\section{The model}
 \label{sec:model}
Consider a cloud of size $L_c$ and of hydrogen density $n_H$ threaded in a magnetic field of intensity $B_0$, oriented along the x-axis. 
Such a one-dimensional configuration is realistic if one considers spatial scales smaller than the magnetic field coherence length in the ISM, i.e. $\sim$ 50-100 pc. Moreover, the magnetic field strength is assumed to be spatially constant and not to change across the transition from outside to inside of the cloud. This is supported by observations \cite{Crutcher10} showing that the magnetic field strength is independent on the density of the interstellar medium as long as the latter remains smaller than $\approx 300$ cm$^{-3}$. The cloud is assumed to be immersed in a diffuse, hot, and fully ionized ISM of density $n_i$. Following \cite{Skill-Strong76}, \cite{Ces-Voelk78}, and \cite{EZ11}, CRs are assumed to propagate along the magnetic field lines only, i.e. diffusion perpendicular to field lines is set to zero.

\begin{figure}
\begin{center}
\includegraphics[width=0.48\textwidth]{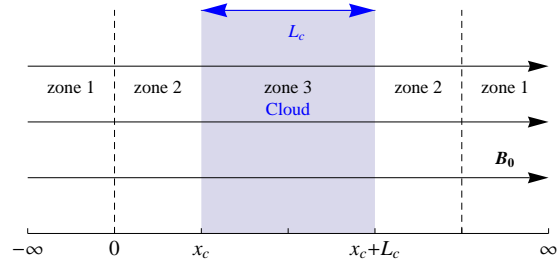}
\end{center}
\caption{Sketch of the simplified 1-D model used to describe the cloud geometry.}
\label{fig:sketch}
\end{figure}

For simplicity, the transition between the low density and ionized ISM and the dense cloud is taken to be sharp and located at $x = x_c$ (see Figure~\ref{fig:sketch}). In order to determine the CR profile along the $x-$axis we solve the one dimensional steady state equation for the transport of CRs in presence of diffusion, advection and energy losses. Written in the rest frame of the plasma the equation reads: 
\begin{equation} \label{eq:fCR}
 \frac{\partial}{\partial x} \left[ D(x,p) \frac{\partial f(x,p)}{\partial x} \right] - v_A(x) \frac{\partial f(x,p)}{\partial x} 
 + \frac{p}{3} \frac{d v_A}{d x} \frac{\partial f}{\partial p} 
 - \frac{1}{p^2} \frac{\partial}{\partial p} \left[ \dot p(x,p) p^2 f(x,p) \right]
 = 0 \,,
\end{equation}
where $f(x,p)$ is the CR distribution function, $D(x,p)$ is the diffusion coefficient, $\dot p(x,p)$ is the momentum losses and $v_A(x)= B(x)^2/\sqrt{4\pi \rho_i}$ is the Alfv\'en speed in the regular magnetic field $B(x)$.  Notice that in this expression we include only the ions mass density, $\rho_i$, because at the wavelengths relevant in this work, the wave frequency is smaller than the charge-exchange frequency between ions and neutrals, hence, while ions oscillate with waves, neutrals have no time to couple with them (e.g. \cite{ZS82}). 
Note that using Eq.~(\ref{eq:fCR}) we assume a diffusive behavior to CRs both outside and inside of the cloud. The validity of such approximation inside the cloud is questionable, because the magnetic turbulence is damped very efficiently by the ion-neutral friction. Nevertheless for the moment we assume that the propagation is diffusive also inside the cloud. We will show that our prediction on the CR spectrum is not strongly affected by this assumption.

Energy losses are presents only inside the cloud (region {\it 3} in Fig.\ref{fig:sketch}). In the energy range between 100 keV and 1 GeV the loss time $\tau_l$ due to ionization of neutral hydrogen can be well approximated by a power law in momentum, which reads:
\begin{equation} \label{eq:t_loss}
  \tau_{l}(p) \equiv \frac{p}{\dot{p}} =  \tau_0 \left( \frac{p}{0.1\, m_p c} \right)^{\alpha}  \left( \frac{n_H}{\rm cm^{-3}} \right)^{-1} \,,
\end{equation}
where the normalization and the slope are $\tau_0 = 1.46 \cdot 10^5$ yr and $\alpha= 2.58$, respectively, and have been obtained fitting the energy losses provided by \cite{Padovani09} (see their Figure 7).

Let us now discuss the boundary conditions. Far from the cloud ($x \ll x_c$) we impose that the CR distribution reduces to the Galactic one $f_0(p)$, 
while at the cloud centre we impose the symmetry condition $\partial_x f(x,p)|_{x=x_c+L_c/2} = 0$. 
The latter condition is different from what has been done in \cite{EZ11},where a condition on the CR gradient far from the cloud, rather than at the cloud center was imposed, to match the CR gradient that one would expect in a Galactic environment. 
However, such a choice may lead to an unphysical solution, because it is what happens inside the cloud that determines the value of the gradient outside of the cloud, and not {\it viceversa}. 
For this reason we decided to use the symmetry condition, which is valid as long as the diffusion approximation holds. In our approach the value of $\partial_x f$ away from the cloud is an outcome of the calculation. 
We also notice that the symmetry condition is valid only for isolated clouds and breaks if the cloud is located near a CR source. Such a situation requires different boundary conditions and will be considered elsewhere.

If we introduce the function $g = D \,{\partial_x f}$, Eq.(~\ref{eq:fCR}) reduces to a linear differential equation of the first order:
\begin{equation} \label{eq:g}
  \partial_x g  - g \, v_A/D + Q =0 \,,
\end{equation}
The nonlinearity of the problem has been hidden in the function $Q(x,p)$ which plays the role of a source/sink term:
\begin{equation} \label{eq:Q}
 Q(x,p) = \frac{p}{3} \frac{\partial v_A}{\partial x} \frac{\partial f}{\partial p} - \frac{1}{p^2} \frac{\partial}{\partial p} \left[ \dot p p^2 f \right] \,.
\end{equation}
The solution for $g$ is:
\begin{equation} \label{eq:sol_g}
 g(x,p) = \int_{x}^{x_c+L_c/2} Q(x',p) \exp\left[-\int_{x}^{x'} \frac{v_A}{D(y,p)} dy \right] dx' \,.
 \end{equation}

We can now write the solution above for the simplified geometry of the MC sketched in Figure~\ref{fig:sketch}. 
Inside the MC we assume a constant density of neutral hydrogen of $n_H= 100$ cm$^{-3}$, with a ionization fraction of $10^{-4}$. For the diffuse ISM we take $n_i=10^{-2}$ cm$^{-3}$, so that the Alfv\'en speed is constant across the transition between the diffuse medium and the MC. 
For the moment we ignore the effect of streaming instability and we assume a Kolmogorov diffusion coefficient outside of the MC: $D(x,p) = D_{kol}(p) \approx 10^{28} (p/mc)^{1/3} \beta$~cm$^2$/s, with $\beta = v_p/c$.
Inside the MC Alfv\'en waves are heavily damped due to ion-neutral friction and the CR diffusion coefficient is $D_c \gg D_{kol}$.  

Under these assumptions it is straightforward to derive from Eq.~(\ref{eq:sol_g}) an expression for $f$ outside of the cloud:
\begin{equation} \label{eq:df_dx}
  f(x,p) = f_0(p) + \frac{1}{v_A} e^{\frac{(x-x_c)}{x_c}} \int_{x_c}^{x_c+L_c/2} Q(x',p) e^{-\frac{v_{A} (x'-x_c)}{D_c}} dx' \,. 
\end{equation}
The exponent outside the integral tells that the CR density outside of the cloud is affected up to a distance of:
\begin{equation} \label{eq:x_c}
  x_c = \frac{D_{kol}}{v_{A}} \approx 300 ~ \beta \left(\frac{B}{5\mu {\rm G}} \right)^{-1} \, \left( \frac{n_i}{0.01\rm cm^{-3}} \right)^{\frac{1}{2}} \, \left(\frac{p}{m_p c} \right)^{\frac{1}{3}} \rm pc \,,
\end{equation}
which can be much larger than the cloud size, and, for a particle energy of $\approx 100$~MeV is of the order of 100 pc, comparable to the magnetic field's coherence length in the interstellar medium.
For distances to the MC larger than $x_c$ the CR distribution reduces to the Galactic one. 
Strictly speaking Eq.~(\ref{eq:x_c}) implies that, for $E \gtrsim 100$ MeV the 1-D approximation breaks down and a more complex transport model should be adopted. However, as we demonstrated in \cite{MorGab15}, when the effect of the streaming instability is taken into account the validity of the 1-D approach is guaranteed up to particle energies of few hundreds of MeV.

Eq.~(\ref{eq:df_dx}) can be further simplified by assuming that the diffusion coefficient inside the cloud is $D_c\gg L_c v_{A}$ (a condition which is easily fulfilled). This implies that the exponent inside the integral reduces to a $\delta$-function. We also note that a critical momentum exists given by the condition that the loss time is longer than the CR propagation time across the cloud $\tau_l(p^*) > L_c^2/2 D_c(p^*)$. Then, for $p \gg p^*$, the CR spatial distribution inside the MC is roughly constant and after some manipulations Eq.~(\ref{eq:df_dx}) becomes:
\begin{equation}
\label{eq:diffstraight}
f(x,p) = f_0(p)-\frac{1}{v_A} e^{\frac{(x-x_c)}{x_c}} \frac{L_c}{2} \frac{1}{p^2} \frac{\partial}{\partial p} \left[ \dot{p} p^2 f_c\right]
\end{equation}
where we also made use of the continuity of the CR distribution function at the MC border $f_c \equiv f(x_c^+) = f(x_c^-)$.

We note that Eq.~(\ref{eq:diffstraight}) can be used to find the solution of the problem also when free streaming of CRs (instead of diffusion) is assumed inside the MC. This is because also under such assumption, a momentum $p^*$ exists above which the CR distribution function can be considered spatially constant. In this case $p^*$ is determined by $\tau_l(p^*) =  L_c/v_{st}$ where $v_{st} \sim v_p/3$ is the free streaming velocity of CRs inside the MC. For $L_c = 10$~pc and $n_H = 100$~cm$^{-3}$ we get $E^* \equiv E(p^*) \approx 0.6$~MeV. The only difference with respect to the diffusive case is that under the assumption of free streaming inside the MC the continuity condition $f(x_c^-) = f(x_c^+)$ has to be considered only as an approximate one.

\begin{figure}
\begin{center}
\includegraphics[width=0.48\textwidth]{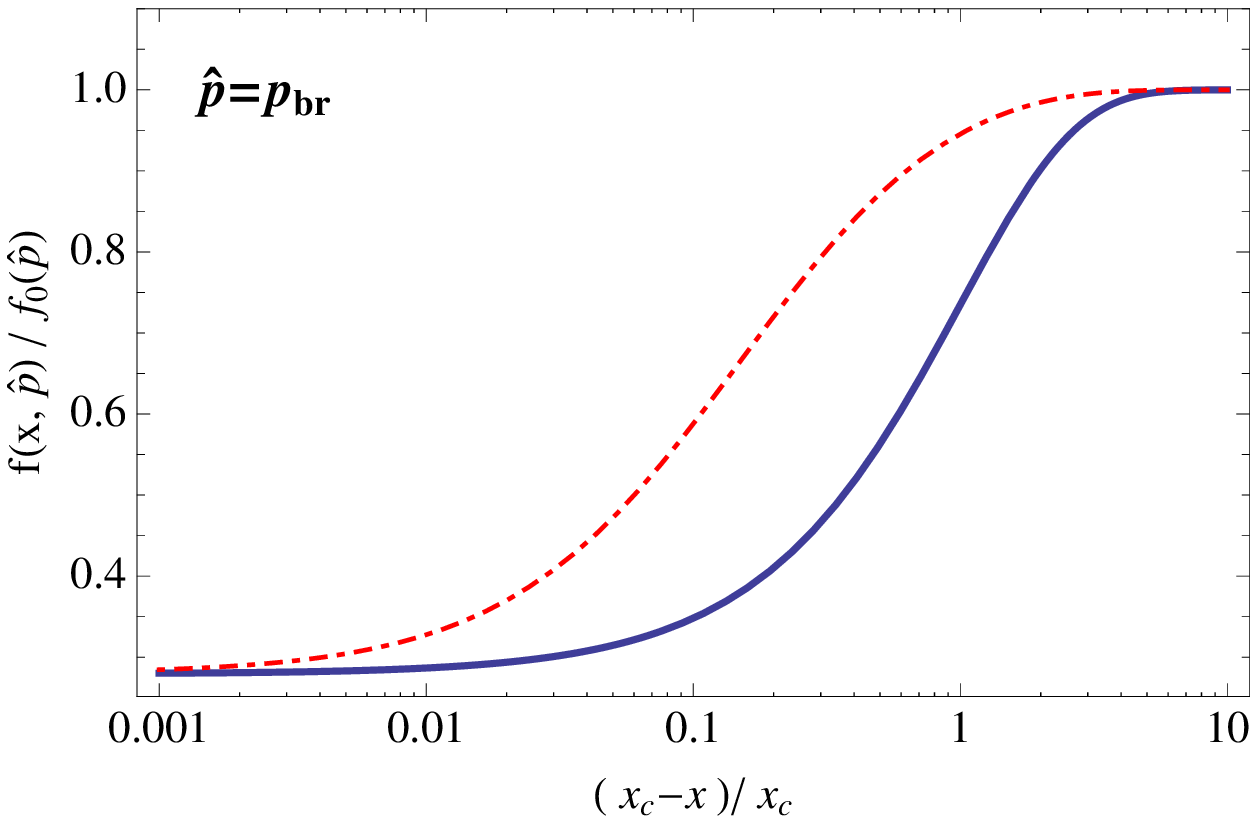}
\includegraphics[width=0.48\textwidth]{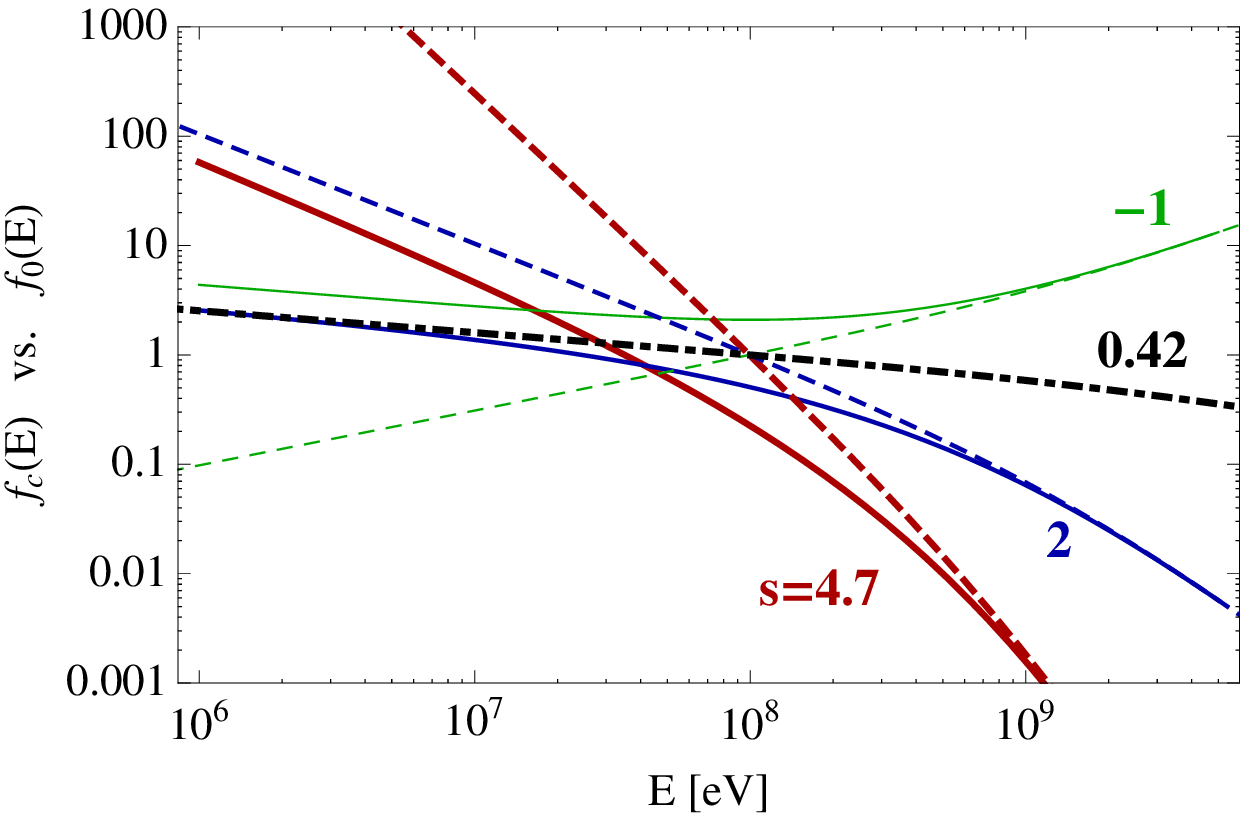}
\end{center}
\caption{Left panel: spatial profile of the CR density outside of the MC according to the solution (2.14) (solid line) for $p=p_{\rm br}\approx 100 \,{\rm MeV}/c$. The dot-dashed line accounts for the presence of streaming instability (see \cite{MorGab15}). Right panel: Spectra of CRs inside the cloud (solid lines) assuming that the spectra far for the cloud are given by power law in momentum of slope $s$ (dashed lines). The dot-dashed line shows the eigenfunction of the problem.}
\label{fig:spectra}
\end{figure}

After using Eq.~(\ref{eq:t_loss}) and performing the derivative in momentum, Eq.~(\ref{eq:diffstraight}) becomes:
\begin{equation} \label{eq:f(x,p)}
  f(x,p) = f_0(p) + \frac{L_c/2}{v_A \tau(p)} \left[ (3-\alpha)f_c+p \partial_p f_c \right]  e^{\frac{(x-x_c)}{x_c}}\,.
\end{equation}
and can be solved for $x = x_c$ to give:
\begin{equation} \label{eq:f1_2}
  f_c(p)= \eta(p) \int_{p}^{p_{\max}} \left(\frac{p'}{p} \right)^3 f_0(p') \,
  		\exp{\left[\frac{\eta(p)-\eta(p')}{\alpha} \right]} \, \frac{dp'}{p'} \,,
\end{equation}
where we used the fact that $\tau$ is a power law. The adimentional function $\eta$ is defined as
\begin{equation}
\label{eq:eta}
\eta(p) \equiv \frac{v_A \tau_l(p)}{L_c/2} \,.
\end{equation}
The condition $\eta(p) = 1$ defines a breaking momentum $p_{\rm br}$, or a breaking energy $E_{\rm br}$ which reads
\begin{equation} \label{eq:E_br}
 E_{br} \simeq 70 \left( \frac{v_A}{100~{\rm km/s}} \right)^{-0.78} \left( \frac{N_H}{3 \times 10^{21}~{\rm cm^{-2}}} \right)^{0.78} {\rm MeV} \,.
\end{equation}
In fact, for $p\gg p_{br}$ (i.e. $\eta(p) \gg 1$), the exponential function in Eq.~(\ref{eq:f1_2}) reduces to  $e^{\eta(p)-\eta(p')} \approx p/\eta(p) \delta(p-p_{\rm br})$. As a consequence $f_c(p) \approx f_0(p)$.
On the other hand for $p\ll p_{br}$ (i.e. $\eta(p) \ll 1$), the exponential function can be approximated as $e^{\eta(p)-\eta(p')} \approx \theta(p-p_{\rm br})$. In this latter case, if we assume a power law for the Galactic CR spectrum, $f_0(p) \propto p^{-s}$, we get
\begin{equation} \label{eq:fc_3}
 f_c(p) \propto \frac{\eta(p)}{(s-3)p^3} \left[  p_{\rm br}^{3-s} - p^{3-s}\right] ~~{\rm for} ~~ p \ll p_{\rm br} \,.
\end{equation}
which implies that, for $p \ll p_{br}$ the solution is a power law $f_c(p) \propto p^{\alpha-3}$ for $s < 3$ and $f_c(p) \propto p^{\alpha-s}$ for $s > 3$. Remarkably, when the CR spectrum in the ISM is a power-law in momentum, $f_0(p) \propto p^{-s}$ with $s = \alpha - 3 \sim 0.42$, the slope of the spectrum of CR is identical inside and outside of the cloud. In other words the function $f_0\propto p^{-0.42}$ is an eigenfunction of the problem, giving $f_c=f_0$. 
The right panel of Figure~\ref{fig:spectra} shows few examples for $f_c(p)$ obtained from Eq.~(\ref{eq:f1_2}) using a simple power law for $f_0\propto p^{-s}$ (showed with dashed lines). 

A remarkable property of solution (\ref{eq:f1_2}) is that $f_c$ does not depend on the diffusion coefficient but only on the Alfv\'en speed in the ISM and on the cloud properties. On the other hand, it is easy to show that the CR spectrum outside the MC does depend on the diffusion coefficient as:
\begin{equation} \label{eq:f1_xp}
 f(x,p) =f_0(p)  + \left[ f_c(p) - f_0(p) \right]  e^{(x-x_c)/x_c} \,.
\end{equation}
which is shown in Figure~\ref{fig:spectra} (left panel) with a solid line, for a spectrum of Galactic CRs $f_0(p) \propto p^{-4.7}$ normalized to an energy density of 1~eV cm$^{-3}$. The typical length scale where the spectrum is affected is $x_c$ as defined in Eq.(\ref{eq:x_c}). As demonstrated in \cite{MorGab15} when the streaming instability is taken into account, the solution inside the cloud, $f_c$, remains unaltered, while the value of $x_c$ my be reduced up to one order of magnitude, depending on the level of damping of magnetic turbulence present in the medium outside the cloud. The left panel of Figure \ref{fig:spectra} also reports the solution when streaming instability is taken into account, but without any damping.

\section{Ionization rate induced by CR} 
\label{sec:ionizaion}
\begin{figure}
\begin{center}
\includegraphics[width=0.47\textwidth]{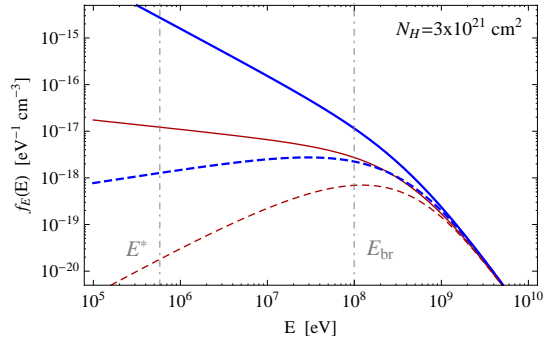}
\end{center}
\caption{Shown are the maximum and minimum Galactic CR proton spectra (solid thick and thin lines, respectively) and the corresponding spectra inside a diffuse cloud with a column density $3\times 10^{21}$ cm$^2$ (dashed thick and thin lines). The vertical dot dashed line mark the critical and the breaking energies $E^*=0.58$ MeV and $E_{\rm br}=99$ MeV.}
\label{fig:spectra2}
\end{figure}

We showed in \S\ref{sec:model} that, as a consequence of propagation, the CR spectrum inside a MC is affected for energies below $E_{\rm br}$ given by Eq.(\ref{eq:E_br}). This fact has very important consequences on the ionization level produced by CRs inside a MC. In fact the ionization cross section peaks for an incident proton energy $\sim 0.1$ MeV, and drops as $E^{-1}$ for $E>E_{\rm peak}$. Hence any process able to modify the CR spectrum in the MeV region will strongly affect the ionization degree of chemicals inside the cloud. The ionization rate is defined as:
\begin{equation} \label{eq:ionization_rate}
  \zeta = 4 \pi \int_{I}^{E_{\max}} j_k(E_k) \sigma_k^{\rm ion}(E_k) dE_k \,,
\end{equation}
where $j_k$ is the CR flux of specie $k$, $\sigma_k^{\rm ion}$ is the ionization cross section and $I$ is the ionization potential of the considered chemical specie. Here we limit our considerations to the ionization rate produced by CR protons on the $H_2$ molecule, $\zeta^{H_2}$, in a typical diffuse cloud. 
We follow the calculation as presented in \cite{julianICRC} which also include the contribution to ionization due to secondary electrons. 
The Galactic CR spectrum to use in Eq.~(\ref{eq:ionization_rate}) is not well know, mainly because the local measured CR spectrum is affected by solar modulation. For this reason, following \cite{Ivlev15}, we adopt a minimum and a maximum spectra which reasonably bound the Galactic spectrum. The lower bound is obtained fitting  the data collected by  the Voyager 1 spacecraft from a region beyond the Solar termination shock, in the energy range between $\sim 5$ MeV and $\sim 50$ MeV. The CR proton flux can be parametrized as
\begin{equation} \label{eq:CR_flux}
  j_{p}(E) = 2.4 \times 10^{15} \, E^\delta \left(E+E_0\right)^{-\beta} \,\rm eV^{-1} cm^{-2} s^{-1} sr^{-1} \,,
\end{equation}
where the minimum spectrum has $\delta= 0.3$ and $\beta=3.0$, while the maximum spectrum has $\delta= -0.5$ and $\beta=2.2$. The crossover energy is $E_0 = 500$ MeV; at $E\gg E_0$, both the spectra have a power low behavior $j_{p}(E)\propto E^{\delta-\beta} = E^{-2.7}$, as detected by the AMS 02 experiment. For the sake of completeness we remember that the flux in energy as reported in Eq.(\ref{eq:CR_flux}) is related to the distribution function in momentum, $f(p)$, by the relation: $j(E)= f_E(E) v(E)/(4\pi) = v(E)p E f(p)/c^2$, where $f_E(E)$ is the distribution function in energy.  
Using Eq.(\ref{eq:CR_flux}) we can estimate the spectrum, $f_c$, inside a cloud. The results are reported in Figure \ref{fig:spectra2} where we compare the energy spectra $f_{E,\min}(E)$ and $f_{E,\max}(E)$ with $f_{c,E,\min}(E)$ and $f_{c,E,\max}(E)$, assuming $L_c= 10$ pc and $n_H= 100$ cm$^{-3}$, which implies a column density of $N_H \simeq 3 \cdot 10^{21}$ cm$^2$. One can see from the figure that already at $E_{\rm br}= 99$ MeV both the spectra are attenuated by a factor $\sim 5$, while at $E^*= 0.58$ MeV, the attenuation reaches roughly 3 orders of magnitudes for both models. For energies below $E^*$ our simplified solution (\ref{eq:f1_2}) provides incorrect prediction.
The ionization rate calculated using the unmodified maximum and minimum spectra are $3.59\times10^{-16}$ s$^{-1}$ and $3.49\times10^{-17}$ s$^{-1}$, respectively. When we account for the propagation, the resulting ionization rate drops to $2.61\times10^{-17}$ s$^{-1}$ for the maximum model and to $1.05\times 10^{-17}$ s$^{-1}$, for minimum one.

It is worth noticing that for typical diffuse clouds with column densities ranging between $10^{21}$ and $10^{22}$ cm$^2$, the measured value of $\zeta^{H_2}$ is between $10^{-16}$ and $10^{-15}$ s$^{-1}$ (see, e.g. \cite{Padovani09}). This implies that CR protons alone cannot account for the whole ionization rate and that the role of electrons has to be carefully considered.

\section{Discussion and conclusions}

In this paper we demonstrated that, due to the balance between advective flux of CRs into the MC and energy losses into the cloud, an equilibrium spectrum forms inside the MC, characterized by a feature at an energy of the order of $E_{br} \approx 100$~MeV for diffuse MCs. Below $E_{br}$ the spectrum falls below (rises above) the CR spectrum in the ISM if the latter is steeper (harder) than $\propto p^{-0.42}$. This fact will have an impact on the estimates of the CR ionization rates in MCs (see \cite{Padovani09} for a review) which are often computed by assuming a ballistic propagation of CRs with a boundary condition $f_c = f_0$ (i.e. the CR spectrum at the border of the cloud is equal to that of interstellar CRs). 
However, before drawing firm conclusions on this issue, the role of CR electrons must be assessed, and this will be done in a forthcoming publication.

The results obtained here are based on the assumption of stationarity. This assumption is justified because the typical time scale of the problem can be estimated from dimensional analysis as $D/v_A^2$ which for a particle energy of 100 MeV gives $\approx 10^6 (D/D_{kol}) (v_A/100~{\rm km/s})^{-2}$~yr, which is always shorter than the dynamical (free-fall) time of the cloud $(G \varrho)^{-1/2} \approx 10^7 (n_H/100~{\rm cm}^{-3})$~yr. 

Finally, we note that a break in the spectrum in the 100 MeV range would not affect significantly the gamma-ray luminosity of the cloud, because the threshold for neutral pion production in proton-proton interactions is at a larger energy, namely $\approx 280$~MeV.

\end{document}